\documentclass[twoside,leqno,twocolumn]{article}

\usepackage{ltexpprt}
\usepackage{hyperref}
\usepackage{booktabs}
\usepackage{tabularx}
\usepackage{graphicx}
\usepackage{algorithm}
\usepackage{algpseudocode}

\begin{document}

\newcommand\relatedversion{}
\renewcommand\relatedversion{\thanks{Thanks to Dr. Eva Tardos at Cornell University for comprehensive feedback, and Ali Abouelatta at Cornell Tech, who helped with some early implementation. \href{https://github.com/ArjunBhalla98/bloom-filter-prims}{Link to code.}}} 

\title{\Large Memory-Efficient Solutions to Large Graph MST problems \relatedversion}
\author{Arjun Bhalla \\ ab2383@cornell.edu}


\date{}

\maketitle


\fancyfoot[R]{\scriptsize{Copyright \textcopyright\ 2023 by SIAM\\
Unauthorized reproduction of this article is prohibited}}




\begin{center}
\begin{abstract}
\begin{changemargin}{0.5cm}{0.5cm} 

\small\baselineskip=9pt Minimum Spanning Trees are a well-studied subset of graph problems. While classical algorithms have existed to solve these problems for decades, new variations and application areas are constantly being discovered. When dealing with large graph problems, however, memory constraints can often be limiting, especially when using these classical methods in memory restricted environments. In this work, we propose an augmentation of Prim's algorithm that can be empirically shown to solve MST problems with a reduction in auxiliary memory usage of over 90\%, and a margin of error of less than 0.2\%. 
\end{changemargin}
\end{abstract}

\end{center}

\section{Introduction}
Finding the Minimum Spanning Tree (MST) of a graph, the set of edges which connects the entire vertex set with minimal cost, is a well-studied category of graph problems. There are a number of core algorithms that are typically used to solve such problems - notably, Kruskal's, Boruvka's, and Prim's Algorithm \cite{ALGOTEXTBOOK}.
\par
Minimum spanning trees have direct applications in network design, clustering gene expression data \cite{WANGETAL}, image registration \cite{SABUNCU}, and image segmentation \cite{BEREG}. Often, these tasks will be quite memory intensive, especially as graph sizes scale, and thus can limit performance or even the problem feasibility. This is especially true in a memory constrained environment (e.g., GPU systems utilising shared memory) or if processing massive graphs \cite{GERAETAL}.
\par
To solve this problem and reduce the memory requirement needed for finding an MST, we propose a novel approach using Bloom Filters. Bloom Filters are a space-efficient randomized data structure for representing a set in order to support membership queries, which allow false positives (importantly, not false negatives). While this is not a desirable trait, the space savings often outweigh this drawback when the probability of an error is controlled \cite{BRODERETAL}. In this paper, we augment Prim's algorithm with Bloom Filters to investigate the difference in memory burden from the original Hash Set based implementation. We compare the memory requirement and error rate of introducing a Bloom Filter in lieu of the Hash Set to track visited nodes, hypothesising that this will lead to a notable decrease in memory consumption, with some acceptable error bound for most applications in the target domains mentioned above.
\section{Preliminaries}
Prior to further discussion, we define the terms used in the paper and discuss the data used.
\subsection{Definitions.}
Many of the definitions used in this paper are the standard vernacular of Computer Science and Graph theory in general. However, for terminology and completion's sake, we define a key term below.

\textbf{Definition 1.1.} A \textit{Minimum Spanning Tree} $\mathcal{T}(E, V)$ on an undirected, edge-weighted graph $G(E,V)$ with an edge set $E$ and vertex set $V$ is a subset of $E$ such that the subgraph $G'(\mathcal{T}, V)$ connects all vertices, forms no cycles, and is of the minimum possible cost to form one giant component with all $v\in V$.

\subsection{Data.}
For initial experimentation, we choose not to use a dataset from a potential application area, such as Image Segmentation or Gene clustering, because the purpose of this paper is to explore a new approach to solving a generic problem. Therefore, we want to test across the most general and unbiased set of data possible. As a result, we generated our graphs using RMAT \cite{RMAT}, specifically using the PaRMAT implementation \cite{PARMAT}. We used the parameters $a=0.45,b=0.22,c=0.22$ to generate a set of undirected graphs with no duplicate or self edges. 

We strongly believe that this is a better approach than choosing a dataset from elsewhere, because a) effectively demonstrating the widespread application of our solution requires generality of the data used instead of domain specific graphs, and b) RMAT is a well known graph generation model with proven results.

\section{Algorithm}
We now present our core contribution and outline the algorithm used.  

\subsection{Overview.} In essence the algorithm is a modification of the widely used implementation of Prim's algorithm. In the typical implementation of this algorithm, a Hash Set is used to keep track of the nodes that the algorithm has visited \cite{ALGORISTS}. However, in our implementation, we replace the Hash Set with a Bloom Filter, a probabilistic data structure introduced to solve set membership problems with limited memory. With just this addition, since we cannot query the Bloom Filter as we can a set, we would at best be able to get an estimate of the cost. However, with the addition of a bitarray to track each edge's membership in the MST (with each edge having a predetermined index in the graph), we can recover the full MST at a fraction of the memory consumed. Thus, this allows us to  drastically reduce space consumed at the cost of a small error margin. It is important to note that in our empirical findings, the margin of error is small ($< 0.2\%$) and that due to the nature of the Bloom Filter the estimated graph will never have duplicate nodes or add edges that would not otherwise be in the deterministic MST. 
\par
It is natural to question the use of a probabilistic data structure like a Bloom Filter over something deterministic, such as another bit array, to store nodes. However, this approach would require having graphs in a specific format, and constraining the actual graph labels and node contents to a small subset of the universe of possible labels, as each node would need to be assigned a unique integer ID. With the Bloom Filter, we can make use of a hash function to allow as many label types as possible, and thus as many applications as possible without requiring pre-processing the graph.

While this algorithm is generalisable, there are certain graph types and graph features that lend themselves to better use cases. For example, this algorithm will likely perform better on dense graphs than sparse graphs. This is because the impact of missing a node (false positive) on a sparse graph is greater, as it is more likely that this node is a bridge node, which could in turn lead to not even exploring a subset of nodes in the rest of the graph. To cover a reasonable set of use cases, we generate graphs such that $|E| \approx O(|V|log|V|)$ in our experiments.

\subsection{Algorithm} 
The algorithm, as presented in Algorithm 1, is a slight modification of Prim's Algorithm. This is a greedy algorithm in which, starting at a random node $s$ in the graph $G$, we first maintain and populate a priority queue $q$ with all of $s$'s neighbours. Then, we traverse the graph in the order given by $q$, adding all new nodes to the visited set, which in this case is maintained as a Bloom Filter $B$. We also add the edge we used to the MST (in this case represented as a bit array $E$, corresponding to each edge). Since we are traversing in the order of the priority queue, we will always be greedily taking the shortest path between nodes, and we will never revisit a node since we check if we have visited this node already prior to processing each node. When the priority queue is empty, we will have a minimum spanning tree on $G$, with a certain cost. We can reconstruct the minimum spanning tree if desired using the edge bit array $E$.
\begin{algorithm}
\caption{Bloom MST}\label{alg:cap}
\begin{algorithmic}
\State $G \gets$ Graph
\State $B \gets$ Empty Bloom Filter
\State $E \gets$ Empty BitArray edgemap
\State $s \gets$ Arbitrary starting node
\State cost $\gets 0$
\State $q \gets$ Empty PQ

\For{n $\in$ s.neighbours()}
\State q.push(n)
\EndFor
\\
\While{q $\neq \emptyset$}:
\State s $:=$ q.pop()
\If{\textbf{not} B.contains(s.SinkNode)}:
\State B.add(s.SinkNode)
\State cost += s.cost
\State E[s.SinkNode.Edge] = 1

\For{t $\in$ G[s.SinkNode]:}
\If{\textbf{not} B.contains(t)}
\State q.push(t)
\EndIf
\EndFor
\EndIf
\EndWhile
\\~\\
\Return{cost, E}

\end{algorithmic}
\end{algorithm}

\subsection{Implementation Details.} The code was implemented in Python 3.7 and used the \texttt{mmh3} library \cite{MMH} for generating universal hash functions for the Bloom Filter and the \texttt{bitarray} library \cite{BITARRAY} for generating bitarrays, as Python does not natively support single-bit bit arrays. We implement the priority queue using the \texttt{heapq} library that is native to Python.

\subsection{Bloom Filter.} The Bloom Filter has a variety of parameters needed to initialise it, all of which could affect experimental results significantly. In order to standardise this and make our experiments widely applicable, we choose them as follows. Given $n$, the size of the input data (graph nodes), and $\epsilon$, the desired error rate (default to 0.01), we calculate $m$, the size of the bucket and $k$, the number of hash functions using the following formulae:
\\~\\
\begin{equation}
m=\frac{-nln\epsilon}{ln2^2},
\end{equation}

\begin{equation}
k=\frac{mln2}{n}.
\end{equation}

\noindent
These formulae are from accepted literature which prove optimality for their respective parameters \cite{STAROBINSKIETAL}.

\subsection{Asymptotic Complexity.}
While the asymptotic running time of Prim's priority queue implementation is $O(|E|log|V|)$ \cite{JOHNSON}, our Algorithm's runtime is $O(k|E|log|V|)$. This is because we query the Bloom Filter each time there is a pop from or add to the heap, and each operation requires computing k hashes. In one iteration, these scale linearly with $|E|$ because we will only consider each edge once.
\par
Practically speaking, however, this is almost negligible for the size of dataset that we are handling, because empirically $k$ is quite low - $\sim 6$ for our experiments. 

\subsection{Error Rate Statistics.}

In order to evaluate our algorithm and assess its effectiveness for a new use case, it would be helpful to have some idea of the number of errors that can be expected given the necessary parameters. As such, we calculate the expectation and the variance of the number of false positives (full hash collisions) for our algorithm.

\subsubsection{Expected Value.} We define $\mathcal{X}\in  \mathcal{N}$ to be a Random Variable representing the number of false positives for an input graph of size $n$, bucket size $m$, and number of hash functions $k$. 

We then define $\mathcal{X}_i \in \{0,1\}$ to be an indicator of whether or not inserting element $i$ will cause a false positive. 
\\~\\
Then, we calculate the expected value of $X$, $\mu_X$, starting summation at $i=2$ as it is impossible for the first insertion to cause a false positive,
\begin{equation}
 \mu_{X}= \sum_{i=2}^n\mu_{X_i}.
\end{equation}
\\~\\
Since $\mathcal{X}_i$ is an indicator variable,
\begin{equation}
 \mu_X = \sum_{i=2}^n \mathcal{P}(\mathcal{X}_i=1).
\end{equation}
\\~\\
We note that $\mathcal{P}(\mathcal{X}_i=1)$ is equivalent to the probability of a false positive given $i-1$ elements currently in the set. This is given by $p_i = (1-e^{-\frac{k(i-1)}{m}})^k$ \cite{BLOOM}. For ease of parsing, we simply refer to this as $p_i$.
\\~\\
Thus, we see that:
\begin{equation}
 \mu_X = \sum_{i=2}^n p_i.
\end{equation}

\subsubsection{Variance.}
We know that the variance for $X$ is given by:
\begin{equation}
 \sigma^2_{\mathcal{X}} = \sum_{i=2}^n \sigma^2_{X_i} + \sum_{i, j, i\neq j}^nCov(X_i,X_j).
\end{equation}
\\~\\
Since the $X_i$ are independent (the probability of a false positive does not get affected by whether or not the last element(s) inserted caused a collision, assuming a uniform distribution for each hash function's outputs):
\begin{equation}
 \sigma^2_{\mathcal{X}} = \sum_{i=2}^n \sigma^2_{X_i}.
\end{equation}
\\~\\
Since $X_i$ is an indicator variable, 
\begin{equation}
\sigma^2_{X_i} = p_i(1-p_i);
\end{equation}
Therefore,
\begin{equation}
 \sigma^2_{\mathcal{X}} = \sum_{i=2}^n p_i(1-p_i).
\end{equation}

\section{Experimental Results}
\subsection{Methodology.}
To measure the impact of our modified Prim's algorithm against the baseline, we track two main metrics:

\begin{itemize}
    \item Memory Consumption
    \item Error Rate (in Bloom Filter implementation)
\end{itemize}
 Originally, we calculated error rate as the difference between the calculated costs of each MST. However, this was not a good metric as the relative magnitude of edges have no bearing on the actual MST, so having an incorrect edge could show a cost difference of 0.01 or 1000. The error metric we then explored was the difference in edge sets for both MST's as a fraction of the total number of edges in the graph, i.e., $\frac{|\mathcal{T}\setminus\mathcal{T}'|}{|E|}$; Here, $\mathcal{T}$ is defined as the deterministic MST, and $\mathcal{T}'$ is the MST generated by our method. This gives more insight into the true number of errors and is invariant under specific edge weights, but it is skewed by the edge density of a given graph. Thus, our final error rate metric compares the delta in the edge sets of the two MSTs against the total edge weights in the baseline implementation's MST, i.e., $\frac{|\mathcal{T}\setminus\mathcal{T}'|}{|\mathcal{T}|}$. This is always guaranteed to lower bound at zero as $|\mathcal{T}'| \leq |\mathcal{T}|$.
\par
For memory consumption, we measure the auxiliary MST representation storage space consumed by both algorithms: for the regular variant, this is the size of the visited (hash)set, and for the Bloom Filter variant, this is the size of the edge-mapped bitarray and the Bloom Filter itself. These results are plotted in Figure 1. We also calculate the total space difference, where we additionally include the space required to store the graph in Figure 2. We have deliberately chosen not to graph the running time, although we note that it displayed no discernible difference on our hardware. This was done because graphing times would not be significant - the algorithms are written in Python, and simple implementation details can cause speeds to vary wildly.  
\par
As explained earlier, we use RMAT to create a generalized data-set to test our algorithms on. The experiments involve running both algorithms on different graphs ranging from 1,000,000 to 10,000,000 nodes in increments of 1,000,000 and measuring the memory consumption and error rate on the baseline and modified Prim's algorithm.

\subsection{Results.}
Results of running experiments on graph sizes from 1,000,000 to 10,000,000 nodes in 1,000,000 node increments. 
\begin{figure}[]
    \begin{minipage}{0.9\linewidth}
        \centering
        \includegraphics[width=\linewidth]{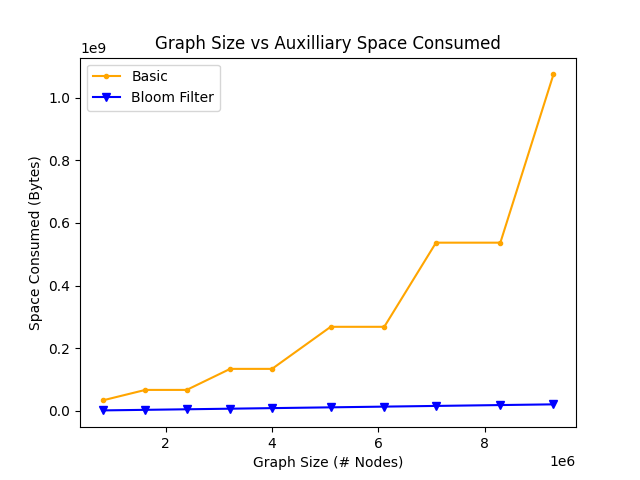}
        \caption{Plot of auxiliary memory consumption (all memory consumption outside of graph storage) in bytes by both the classical and the Bloom Filter implementations of Prim's Algorithm.}
    \end{minipage}%
    \\
    \begin{minipage}{0.9\linewidth}
        \centering
        \includegraphics[width=\linewidth]{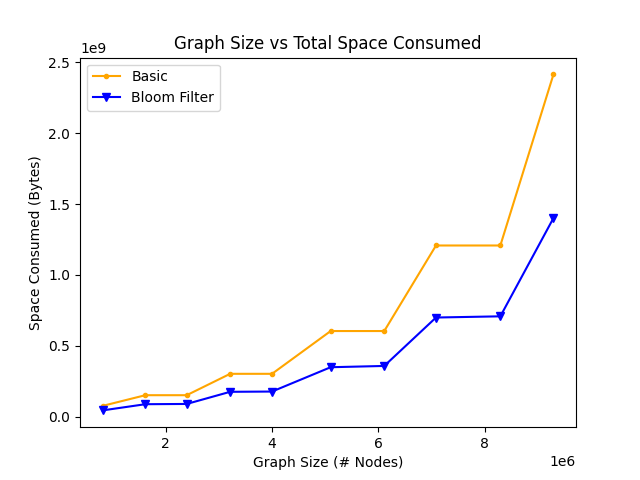}
        \caption{Plot of total memory consumption in bytes by both the classical and the Bloom Filter implementations of Prim's Algorithm.}
    \end{minipage}\par\vspace{3pt}
    \begin{minipage}{0.9\linewidth}
        \centering
        \includegraphics[width=\linewidth]{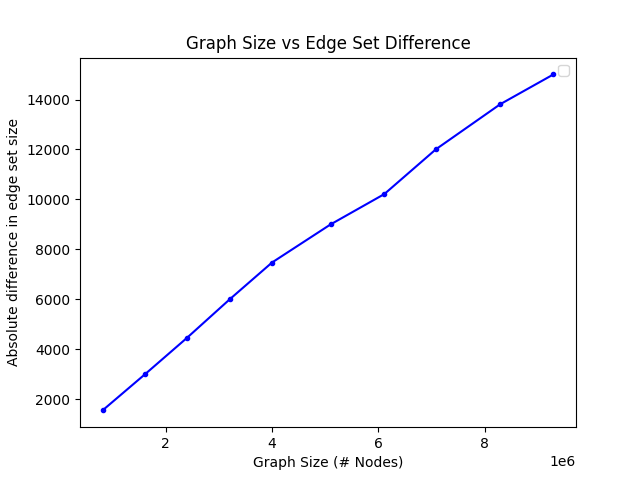}
        \caption{Plot of the measured error of the Bloom Filter implementation.}
    \end{minipage}
\end{figure}  
\begin{table*}[]
\begin{minipage}{\linewidth}
        \centering
        \caption{Table of results from experiments, including expected value ($\mu_X$) and standard deviation ($\sigma_X$)}
        \small
        \begin{tabularx}{\textwidth}{@{} l *{8}{c}@{}}
            \toprule
            \# Nodes & Baseline Space (Bytes) & BF Space (Bytes) & Difference & \# Incorrect Edges & $\mu_X$ & $\sigma_X$ & Error \\
            \bottomrule 
            \
            814,817 &  33,554,648 & 1,728,455 & 94.85\% & 1553 & 1356.38 & 36.73 & 0.19\%\\
            \ 
            1,619,565 & 67,109,080 & 3,519,847 & 94.76\% & 3013 & 2696.10 & 51.78 & 0.19\%\\
            \ 
            2,396,625 & 67,109,080 & 5,306,319 & 92.10\% & 4448 & 3989.55 & 62.99 & 0.19\%\\ 
            \  
            3,212,885 & 134,217,944 & 7,157,922 & 94.71\% & 6016 & 5348.34 & 72.93 & 0.19\%\\
            \ 
            4,000,236 & 134,217,944 & 8,988,600 & 93.33\% & 7474 & 6659.00 & 81.37 & 0.19\%\\
            \ 
            5,101,912 & 268,435,888 & 11,499,320 & 95.72\% & 9,001 & 8492.91 & 91.90 & 0.18\%\\
            \ 
            6,109,280 & 268,435,888 & 13,795,112 & 94.86\% & 10,212 & 10169.83 & 100.56 & 0.17\%\\
            \ 
            7,081,825 & 536,871,776 & 16,011,542 & 97.07\% & 12,010 & 11788.78 & 108.27 & 0.17\%\\
            \ 
            8,291,108 & 536,871,776 & 18,767,498 & 96.50\% & 13,814 & 13801.82 & 117.15 & 0.17\%\\
            \ 
            9,289,165 & 1,073,743,552 & 21,042,070 & 98.04\% & 15,005 & 15463.24 & 124.00 & 0.16\%\\
            \toprule 
            Average & N/A & N/A & 95.19\% & N/A & N/A & N/A & 0.18\%\\
            \bottomrule
        \end{tabularx}
    \end{minipage}
\end{table*}
\subsubsection{Key Takeaways}
\begin{itemize}
    \item Achieved an average \textbf{95.19\%} decrease in memory usage for visited set computation with our method.
    \item This was with a \textbf{0.18\%} average error rate - that is, the Bloom Filter implementation missed just fewer than 1/500 edges in each graph. 
\end{itemize}

\subsubsection{Memory Consumption.}
As we see in Figure 1 and Table 1, there is a stark difference in the memory consumed by both of the methods - our implementation is almost two orders of magnitudes more memory efficient. The Bloom Filter graph follows a shallow but constant linear trend.
\par
If comparing the total memory consumed, as in Figure 2, which includes the space required to store the graph, we see a more modest but still impressive space reduction. Here, our approach, regardless of graph size, uses about $57.8\%$ of the total memory of the classical Prim's Algorithm.
\par
The rather erratic graph for the baseline implementation in Figure 1 and Figure 2 can be explained simply by the way in which hashsets work - the $O(1)$ lookup / addition is amortized because when a certain number of elements are stored in the set, to prevent excess chaining and thus longer lookup / addition times, the set dynamically increases its size to meet new demand.
\par
We can thus infer that there is what appears to be a shallow but strong linear relationship between graph size and space consumed by our method. This suggests that these results directly support our hypothesis and are unlikely to be caused by some error or oversight. 

\subsubsection{Error Rate.}

Again, this seems to be fairly linear with the size of the graph as we see in Figure 3, which is consistent with the mathematics in Section 3, but the perturbation of some data points with respect to a linear trendline serves as a further reminder that the error percentage that was calculated is not a guarantee, but rather an empirical finding that supports a linear trend. 
\par
From this, we see that error scales with the number of nodes in the graph as opposed to the number of edges.

\section{Application to Image Segmentation}

Now that an experimental baseline has been established alongside limitations and expectations of the method, we attempt a visualisation of the results of our algorithm in a target application area, comparing it with the results of running the baseline method. We choose this application area due to its straightforward nature, and the fact that image processing tasks can be notoriously memory-intensive, especially as size and quality increase \cite{HAJALIETAL}

\subsection{Method.}
We use a simple MST-clustering based image segmentation algorithm, based on commonly used methods and the implementation for the SciKit-Learn Python Library \cite{MSTCLUSTERING}. Despite there existing more sophisticated methods to perform this task  \cite{LVETAL}, \cite{CHAETAL}, we use this straightforward approach to keep the focus on the our algorithm and its impact instead of modifying another approach to this domain-specific problem.
\par
Given an MST and a cost threshold, we first trim all edges in the tree above the threshold, then perform a Breadth-First search on each of the consequently formed sub-components to assign them the same label. Each node here is a pixel, and the cost of each edge in this case is equivalent to the square of the Euclidean distance between the two nodes' RGB values. We then segment using a cost threshold of 100, chosen after some experimental runs to provide the most consistent segmentation across different images. We determine this threshold after running through a set of 30 images (10\% of the dataset), and looking at 2 factors. The first is the number of clusters formed at different thresholds and the volatility of each segmentation (how the number of clusters changed with small perturbations in the threshold). The second is a manual inspection of segmentation quality, and a look at the overall coherency of the segmentation.
\par
The images that we use are from a standard dataset, the Berkeley Image Segmentation Dataset (BSDS300) \cite{BSDS}. We set destination edges for each node to its immediate neighbours - 8 possibilities in total, with up, down, left, right, and all diagonals. All images in this set are of size 321x481 or 481x321, each resulting in a graph with 154,401 nodes and 1,230,400 edges.

\subsection{Example Results.}
\begin{figure}
    \centering
    \includegraphics[width=0.3\linewidth]{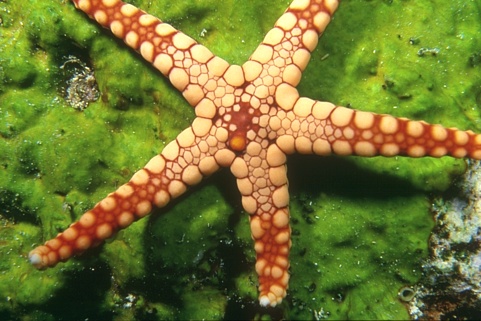}
    \includegraphics[width=0.3\linewidth]{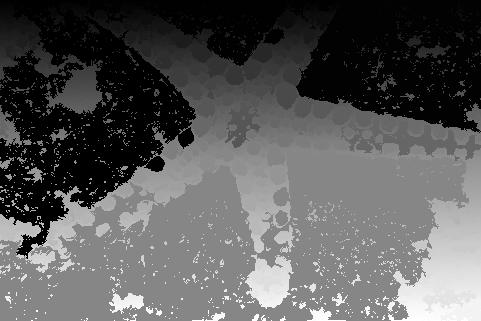}
    \includegraphics[width=0.3\linewidth]{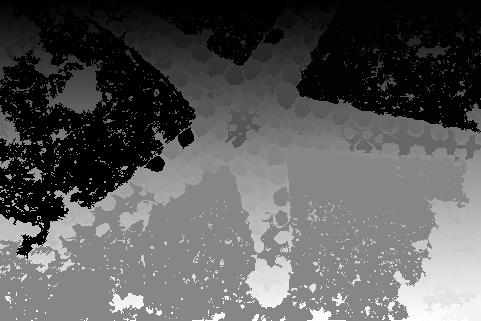}
    \\~\\
    \includegraphics[width=0.3\linewidth]{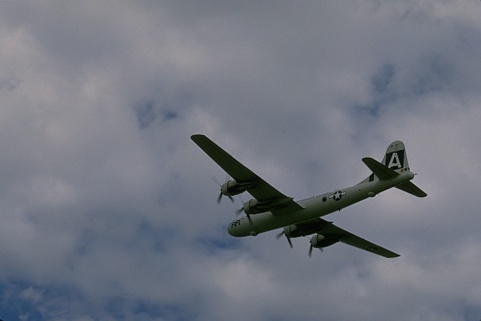}
    \includegraphics[width=0.3\linewidth]{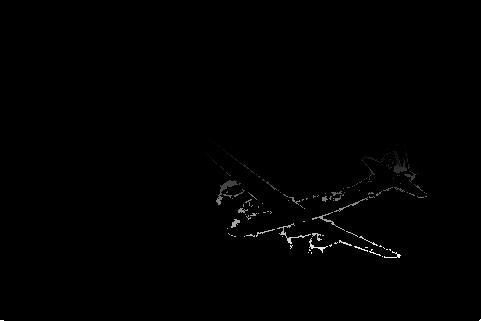}
    \includegraphics[width=0.3\linewidth]{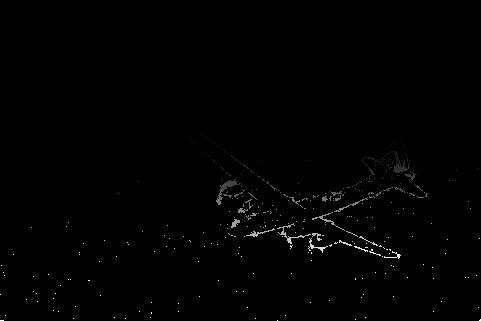}
    \\~\\
    \includegraphics[width=0.3\linewidth]{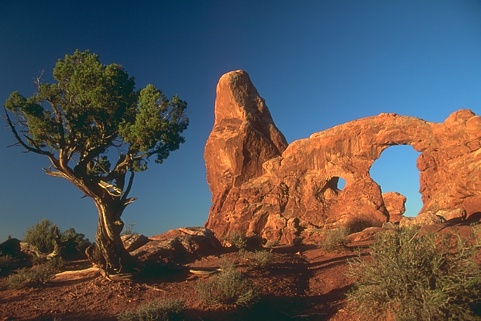}
    \includegraphics[width=0.3\linewidth]{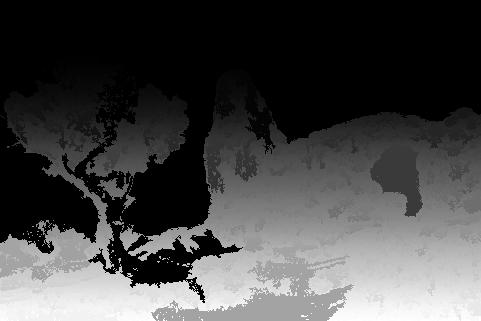}
    \includegraphics[width=0.3\linewidth]{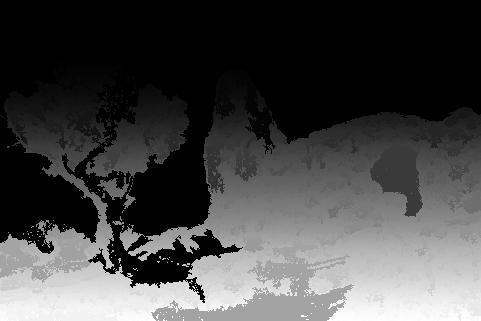}
    \caption{The original image, Prim's Algorithm MST segmentation, and our Algorithm's segmentation.}
    \label{fig:example_segment}
\end{figure}

The results of running segmentation and averaging across the provided training set split of the BSDS 300 (201 images) yielded an average difference of \textbf{$\sim$1.229\%} in the number of clusters formed by both algorithms. However, at this thresholding level, the number of clusters formed is reasonably high, on the order of magnitude of $10^3$, so functionally the clusters representing major elements are preserved (see \ref{fig:example_segment}). The high cluster count is more a symptom of this approach to image segmentation as opposed to our algorithm.

\section{Conclusion}

We introduce a new method for solving MST problems that uses orders of magnitude less memory than Prim's algorithm, a traditional and very popular baseline, while introducing a negligible slowdown and a negligible (application-dependent) empirical error rate. This will always result in a cost less than or equal to the true cost, provided all edge weights are non-negative. This method is able to solve large graph MST problems in memory constrained situations (e.g., GPU graph computations \cite{GERAETAL}), and where producing an approximate result is acceptable.
\par
The direct implications of this are that solving MST problems on massive graphs will be much more accessible commercially, not requiring large amounts of RAM or specialised hardware, as well as providing another approach for constrained memory GPU based graph processing problems. This could, depending on the results of future work, mean that another avenue of approaching problems in application areas such as fast image segmentation or image registration on massive datasets would be available to those with hardware constraints, or that GPUs would be able to more efficiently solve large MST problems. For example, a dataset that would take 100GB of memory to process using Prim's algorithm could be approximately solved with only $\sim$ 6.47GB using this method.

\section{Future Work}
We believe that this is a promising beginning to what will hopefully be a fruitful and useful application of Bloom Filters. However, there is still much that is left unexplored. We hope to try further domain application to test the limits of this approach, and to realistically attempt experimentation with solving graph problems on GPUs. To do this it would also behoove us to build a robust implementation of this in a lower level language, such as C or C++, such that we could properly assess performance as well.
\par
We could also explore changing variables in our initial approach. As mentioned earlier, we have implemented this method with the assumption that we know the full dataset size ahead of time. However, what would happen if we relaxed this assumption and had an unseen graph? Is it useful to manually set the $m$ and $k$ parameters for different application areas? What are the error bounds on an estimate of the graph size that would still render this method effective? These questions can certainly be the basis for interesting future research directions.

\end{document}